\documentstyle[12pt,fleqn]{article}
\begin{document}
\baselineskip= 22 truept
\def\be{\begin{equation}}
\def\ee{\end{equation}}
\def\bea{\begin{eqnarray}}
\def\eea{\end{eqnarray}}
\def\pa{\partial}
\begin{titlepage}
\begin{flushright}
IP/BBSR/96-04\\
hep-th/9601067 \\
\end{flushright}
\vspace{1cm}\begin{center} {\large \bf Orientifold 
and Type II Dual Pairs}\\ \vspace{1cm} {\bf 
Alok Kumar 
 }\\ Institute of Physics, \\Bhubaneswar 751 005,
INDIA \\ email: kumar@iopb.ernet.in \\ 
\today
\end{center} 
\thispagestyle{empty}
\vskip 4cm
\begin{abstract}

In this paper we present a symmetry of 
a toroidally compactified type II 
string theory. This symmetry has the 
interpretaion that it interchanges the left and the 
right-moving worldsheet coordinates and reverses the 
orientations of  some of the spatial coordinates.
We also identify another discrete symmetry of the 
type II theory  which is related to 
the above one by a nontrivial U-duality element of 
string theory. This 
symmetry, however, has trivial action on the worldsheet 
coordinates and corresponds to an improper T-duality rotation. 
We then construct examples of 
type II dual pairs in four dimensions by
modding out the known type II dual pairs 
by the above symmetries. 
We show the explicit matching of the spectrum 
and supersymmetries in these examples.

\end{abstract}
\vfil
\end{titlepage}
\eject

String-String duality conjecture has acquired a great deal of
support from a number of independent checks performed 
in last one year. This conjecture implies that there are pairs
of string theories where strong coupling limit of one 
can be interpreted as the weak coupling limit of another
and vice versa.
Earliest support \cite{duff,hull,witten}
in favour of this conjecture have been the 
matching of the neutral particle spectrum and mapping of the 
the string effective action for the heterotic-type IIA pair of 
strings in six dimension for the case when the heterotic string is 
compactified on $T^4$ and the type IIA is compactified on
$K3$\cite{hull,witten,townsend}. This conjecture 
received further support from the demonstration that 
the fundamental string solutions of one theory is mapped to
the solitons of another one by a strong-weak coupling 
duality\cite{duff,sen1}. 
The emergence of chiral worldsheet action of the 
heterotic string theory\cite{harvey,ferrara}, 
with the right number of fermionic
and bosonic zero modes as a soliton worldsheet action 
of the type IIA theory for K3 compactification, was another 
evidence of the existence of this duality symmetry. Some of
these results have been checked by a direct calculation of
the string loop amplitudes for the heterotic string theory and
comparing them with the exact results of the 
type IIA theory\cite{kachru,anton}. 

Starting from the above heterotic-type II dual pair,
many other nontrivial examples of the dual pairs of the 
heterotic-type II string theory were constructed in
\cite{vawit,scsen,chaudh,quev,gao}. 
Many of these constructions are based on 
an {\it adiabatic} argument which implies that the 
two theories which are ``strong-weak" dual of one another 
remain so during the process of 
compactifications on circles of arbitrary radii.
This is also true in the case when the 
two theories in higher dimensions are modded out by 
a pair of discrete symmetries, related by
stong-weak coupling duality,  
as long as these symmetries 
have no fixed points. An easy way to avoid fixed 
points is to compactify on circles
of arbitrarily radii and combine the discrete 
operations on the individual 
higher dimensional theories 
with the translations on these circles. 
In these cases, 
the twisted sectors are all massive and the left-right 
matching condition of the spectrum is maintained.

The arguments of the above paragraph were applied  
by Sen and Vafa\cite{seva} to obtain the dual 
pairs of type II theories in four as well as in six 
dimensions. In this case they explicitly 
identified a matrix transformation  which represents the 
strong-weak coupling duality operation. This gives a dual 
pair, for every pair of T-duality symmetries
in six dimensions which are used to 
mod out the original theories, when they are compactified 
down to four dimensions. A method for 
constructing dual pairs by going beyond the {\it adiabatic}
arguments was also given in \cite{seva}. 
This method was also applied to construct many
four and six dimensional examples in
\cite{seva}. 

In this paper we obtain 
a new $Z_2$ symmetry, referred as $Z_2^o$, of the 
type II effective action. The transformations of 
space-time fields under this $Z_2$ can be 
identified as an orientation reversing symmetry 
of the worldsheet 
together with the reversal of orientations 
of some of the spatial coordinates.
This symmetry is similar in its action to the 
orientifold examples considered in the case of
$K3$ compactification\cite{vawit}. 

We also identify another $Z_{2}$ symmetry,
referred as $Z_{2}^{\star}$, of the type II 
action which is related to $Z_2^o$ in the 
previous paragraph by the operation of 
strong-weak coupling duality. 
Unlike $Z_{2}^o$, this is a part of the 
T-duality symmetry of string theory. However,
it  lies in the set of T-duality rotations 
which are not connected to the identity 
continuously. We then work out 
examples where we obtain new type II 
dual pairs of string theories by modding out the 
known ones by the above discrete symmetries .
In physical terms, the modding out by 
$Z_{2}^o$ in general corresponds to the construction of 
string theory on a non-orientable 
worldsheet. Our method therefore gives examples of
dual pairs where underlying worlsheet theory 
in one side is defined on  orientable 
Riemann surfaces and in another side 
it is  defined on non-orientable ones\cite{vawit}.

We now start by writing down the type IIA string 
effective action in ten dimensions and identifying 
the $Z_2^o$ symmetry descibed in the introduction. 
The type IIA action is given as\cite{harvey}
\bea
S = {1\over {16 \pi {\alpha'}^4}}\int d^{10}x 
        \sqrt{-g}\left[  e^{-2\phi} ( R  + 
4 {(\nabla \phi)}^2 - {1\over 3} H^2 )
-\alpha' G^2 - {\alpha'\over {12}}F'^2 \right.\nonumber \\ 
\left. -{\alpha'\over {288}} \epsilon^{M_1 ...M_{10}}
F_{M_{1}...M_{4}}F_{M_{5}...M_{8}} B_{M_{9} M_{10}} \right],
                        \label{action}
\eea
where $g$ is the ten dimesnional string
metric and (M,N = 0, .., 9).   
$G_{M N}$ is the field strength 
for the RR gauge field $A_M$ in ten dimensions:
$G_{M N}=2 \pa_{[M} A_{N]}$, 
$H_{M N P}$ is the field
strength for the 2-Form field
 $B_{M N}: H_{M N P}=3 \pa_{[M} B_{N P]}$ and 
$F_{M N P Q} \equiv 4 \pa_{[M} C_{N P Q]}$ is the
field strength for a 3-Form field $C_{M N P}$
and $F'_{MNPQ} = F_{MNPQ} + 8 A_{[M} H_{NPQ]}$. 
A six dimensional $N=4$ supersymmetric string
theory can be obtained from the above
action by a compactification on a four 
torus. This essentially implies the 
independence of the background fields in the above action 
with respect to the coordinates $x^6, .. ,x^9$. 
To discuss the symmetries we are interested in, we 
put further constraint that the backgrounds are 
also independent of the $x^5$ coordinate.
As a result string theory under consideration is the type IIA
theory compactified on $T^4\times S^1$.

We now show that the type IIA action with these restrictions
is invariant under a 
new discrete transformation whose action is a combination of 
orientation reversing transformations on space-time and 
the string worldsheet. Since such a symmetry acts by 
changing the sign of the four dimesional antisymmetric 
tensor, it is not part of any T-duality transformation.
On the other hand, we will see that, the strong-weak 
duality transforms it to another symmetry of the type II
action which can be interpreted as an improper 
T-duality rotation. 

This symmetry, $Z_2^o$, is now found 
by a direct investigation of the terms in action (\ref{action}) by 
decomposing the ten dimensional indices $M$ as 
$M \equiv (\mu, 5, i)$.  We mention that 
various indices in this paper run as:
$M = 0,..,9$, $\mu = 0, ..,4$, $\bar{\mu} = 0,..,5$, 
$\hat{\mu} = 0,..,3$, $i = 6, .., 9$ and 
$m= 6,..,8$. It can then be 
shown that the action (\ref{action}) is invariant 
under the following set of transformations 
acting on the 
components of the ten dimensional fields as:

\bea
\pmatrix{ g_{\mu \nu}  & g_{\mu 5} &g_{\mu j}\cr
           g_{5 \nu}    & g_{5 5}   &g_{ 5 j}\cr
           g_{i \nu}    & g_{i 5}   &g_{ i j}}
           \rightarrow
\pmatrix{ g_{\mu \nu}  & -g_{\mu 5} & -g_{\mu j}\cr
         -g_{5 \nu}    & g_{5 5}     & g_{5 j}\cr
        - g_{i \nu}    & g_{i 5}     & g_{i j}},\nonumber\\
\pmatrix{ B_{\mu \nu} & B_{\mu 5 } &  B_{\mu j}\cr
          B_{5 \nu}    &  0         &   B_{5 j}\cr
          B_{i  \nu}   &  B_{i 5}   &    B_{i j}}  
                         \rightarrow
\pmatrix{ -B_{\mu \nu}  & B_{\mu 5} & B_{\mu j}\cr 
           B_{5 \nu}   & 0         & - B_{5 j}  \cr
           B_{i \nu}   & -B_{i 5}  &  -B_{i j}},\nonumber\\
\pmatrix{ A_{\mu} \cr
          A_{5} \cr
          A_{i} } 
\rightarrow
\pmatrix{  A_{\mu} \cr
          -A_{5} \cr
          -A_{i}},
\pmatrix{ F_{\mu_{1}  . .  \mu_{4}} \cr
          F_{\mu_{1}  . \mu_{3}  i} \cr
          F_{\mu_{1}  \mu_{2} i j}   \cr
          F_{\mu_{1}  i j k }  \cr
          F_{\mu_{1}  . \mu_{3} 5}  \cr
          F_{\mu_{1}  \mu_{2} j 5} \cr
          F_{\mu_{1}  i j 5}}
             \rightarrow
\pmatrix{ -F_{\mu_{1}  . . \mu_{4}} \cr
           F_{\mu_{1}  . \mu_{3} i} \cr
          -F_{\mu_{1}  \mu_{2} i j} \cr
           F_{\mu_{1}  i j k}       \cr
           F_{\mu_{1} . \mu_{3} 5}  \cr
          -F_{\mu_{1}  \mu_{2} j 5} \cr
           F_{\mu_{1} i j 5}}.  
                \label{z2o}
\eea\\

The above symmetry 
transformations on the space-time fields have 
a simple worldsheet interpretation.
Since this symmetry takes $B_{\mu \nu} 
\rightarrow - B_{\mu \nu}$ and $B_{i j} \rightarrow
- B_{i j}$, one has to interchange $(z \leftrightarrow
\bar{z})$ to keep the worldsheet action invariant.
However, since $B_{\mu 5}$ remains invariant and 
$B_{i 5}$ changes its sign, the $z\leftrightarrow \bar{z}$ 
transformation on the worldsheet has 
to be accompanied by transformations $X^5 \rightarrow
-X^5$ and $X^i \rightarrow -X^i$. The transformation 
properties of $g_{MN}$, $A_M$ and $C_{MNP}$ 
further support this observation.
This feature clearly distinguishes $Z_2^o$ symmetry 
from a T-duality symmetry, which has a trivial action on 
the string worldsheet coordinates. These 
properties have also been utilized earlier in  
understanding duality aspects of 
open string theory\cite{horava,leigh}. 
In this case, open string theory emerges as a 
twisted sector of the closed string theory\cite{horava}, 
when the former ones
are modded out by the symmetries of the above type. 
This aspect has also been studied recently 
in several papers\cite{mukhi}. This 
leads to a model which is a combination of space-time and 
worldsheet orbifolds. 
In our case, however, above symmetry will be combined with 
another one to avoid any fixed points. Then
twisted sectors are all massive by construction. The
interpretation of modding out by the above symmetry is 
to describe a string theory on non-orientable 
Riemann surfaces. 
The nature of the symmetry discussed above is  
similar to another one for the type IIA strings
compactified on $K3$, as described in \cite{vawit}. This symmetry
can also be obtained on a similar line by studying the 
corresponding type IIA action \cite{sen1}. 

One can also use arguments along the lines of 
\cite{vawit,seva} to 
find an operation on fields which is related to those in 
equations (\ref{z2o}) by strong-weak duality transformation, 
namely the action of an element of $U$-duality, $U_0$.
In the present case, we first write down a symmetry 
transforamtion below and later on, using the results of 
\cite{vawit} and \cite{seva}, 
argue that it is related to the $Z_2^o$ symmetry 
of equation (\ref{z2o}), by the $U$-duality element, $U_0$.
This symmetry $Z_2^*$ has the form:

\bea
\pmatrix{  g_{\mu \nu} & g_{\mu 5} & g_{\mu m} & g_{\mu 9} \cr
                 g_{ 5 \nu} & g_{ 5 5} & g_{ 5 m} & g_{5 9}  \cr
                 g_{ m \nu} & g_{m 5}  & g_{m n} & g_{m 9}  \cr
                 g_{ 9 \nu} & g_{9 5}  & g_{ 9 m} & g_{9 9} }
\longrightarrow
\pmatrix{ g_{\mu \nu} & -g_{\mu 5} & -g_{\mu m} & g_{\mu 9} \cr
         -g_{ 5 \nu}  & g_{ 5 5}   &  g_{ 5 m}  & -g_{ 5 9} \cr
         -g_{ m \nu}  & g_{m 5}    &  g_{m n}  & -g_{m 9}  \cr
          g_{ 9 \nu} & -g_{9 5} & -g_{ 9 m} & g_{9 9} },\nonumber\\
\pmatrix{ B_{\mu \nu}& B_{\mu 5}& B_{\mu m}& B_{\mu 9} \cr      
          B_{5 \nu} & 0 & B_{5 m} & B_{5 9} \cr
          B_{m \nu} & B_{m 5} & B_{m n} & B_{m 9} \cr  
          B_{9 \nu} & B_{9 5} & B_{9 m} &  0 }
\longrightarrow
\pmatrix{ B_{\mu \nu} &-B_{\mu 5} &-B_{\mu m} &B_{\mu 9} \cr
         -B_{5 \nu} &0   &B_{5 m} & -B_{5 9}\cr
         -B_{m \nu} & B_{m 5} & B_{m n} & -B_{m 9}\cr     
          B_{9 \nu} & -B_{9 5}& -B_{9 m} & 0 } ,\nonumber\\
\pmatrix{ A_{\mu} \cr 
          A_{5} \cr 
          A_{m} \cr                
          A_{9}}
\longrightarrow
\pmatrix{ A_{\mu} \cr 
        -A_{5} \cr 
        -A_{m} \cr  
         A_{9}},
\pmatrix{ F_{\mu_{1} \mu_{2} \mu_{3} \mu_{4}}\cr
         F_{\mu_{1} \mu_{2} \mu_{3} m}\cr
         F_{\mu_{1} \mu_{2} m n}\cr
         F_{\mu_{1}  m n p}\cr
         F_{\mu_{1} . \mu_{3} 5}\cr
         F_{\mu_{1} \mu_{2} m 5}\cr
         F_{\mu_{1} m n 5}\cr
         F_{\mu_{1} . \mu_{3} 9}\cr
         F_{\mu_{1} \mu_{2} m 9}\cr
         F_{\mu_{1} m n 9} \cr
         F_{\mu_{1} \mu_{2} 5 9}\cr
         F_{\mu_{1} m 5 9}}
\longrightarrow
\pmatrix{ F_{\mu_{1} . . \mu_{4}}\cr
         -F_{\mu_{1} . \mu_{2} m}\cr
          F_{\mu_{1} \mu_{2} m n}\cr
         -F_{\mu_{1} m n p}\cr
         -F_{\mu_{1} \mu_{2} \mu_{3} 5}\cr
          F_{\mu_{1} \mu_{2} m 5}\cr
         -F_{\mu_{1} m n 5}\cr
          F_{\mu_{1} . \mu_{3} 9} \cr
         -F_{\mu_{1} \mu_{2} m 9}\cr
          F_{\mu_{1} m n 9}\cr
         -F_{\mu_{1} \mu_{2} 5 9}\cr
          F_{\mu_{1} m 5 9}}.
                \label{z2*}
\eea

To interpret this symmetry, we note that it leaves both 
$B_{\mu \nu}$ and $g_{\mu \nu}$ components 
invariant, a property of the 
T-duality symmetry. In fact, we will see later on that 
this symmetry transformation corresponds to a 
T-duality rotation in six dimensions, together with a 
discrete coordinate transformation $x^5 \rightarrow 
- x^5$. Moreover, these are the improper T-duality 
rotation, not connected to identity continuously. We also note
that the T-duality rotations used in \cite{seva} 
are all proper ones,
since they all belong to the set $SO(5) \times SO(5)$. 
Our results therefore show that the discrete symmetries outside 
those of the set considered in \cite{seva} 
can be used to construct new dual pairs of string theories. 

Now, to identify the symmetries in (\ref{z2o}) and 
(\ref{z2*}) as pairs of
symmetries related by the strong-weak duality element $U_0$, we
combine the set of fields in these equations in the form:

\be
\tilde{A}_{\bar{\mu}} \sim \left(\begin{array}{cc} 
                                g_{\bar{\mu} 6} \\
                                 .      \\
                                 .      \\
                                g_{\bar{\mu} 9} \\
                               B_{\bar{\mu} 6} \\
                                .       \\
                                .       \\
                                B_{\bar{\mu} 9}
        \end{array}     \right), 
\tilde{\psi} \sim\left(\begin{array}{cc} A_{6} \\
                              .     \\
                              .     \\
                              .     \\ 
                             A_{9}  \\
                             C_{7 8 9} \\
                             C_{6 8 9} \\
                             C_{7 6 9} \\
                             C_{6 7 8} 
              \end{array}    \right),
\tilde{K}_{\bar{\mu}} \sim \left(\begin{array}{c}   
                                        C_{\bar{\mu} 6 9}  \\               
                                     C_{\bar{\mu} 7 9}  \\
                                     C_{\bar{\mu} 8 9}   \\
                                    \tilde{K}_{\bar{\mu}}         \\
                                     C_{\bar{\mu} 6 7}   \\
                                     C_{\bar{\mu} 6 8}  \\
                                     C_{\bar{\mu} 7 8}   \\
                                     A_{\bar{\mu}}
                   \end{array}   \right), 
        \label{defns}
\ee
where $\bar{\mu}$, as mentioned before, is a six dimensional 
index and $\tilde{K}_{\bar{\mu}}$ is the field 
dual to $C_{\bar{\mu_1}..\bar{\mu_3}}$ in six dimensions. 
The internal components of the metric $g_{ij}$ and the
antisymmetric tensor $B_{i j}$ are arranged as:
\be
\tilde{M} = \left(\begin{array}{cc}  g^{-1}   &       -g^{-1}B   \\
                             Bg^{-1}    &   g- Bg^{-1}B 
               \end{array} \right).
                        \label{defns1}
\ee

To make connection with the results in \cite{seva}, 
we note that in \cite{seva}
the set of fields in six dimensions 
were arranged in the following form:
(i) Graviton, dilaton
and the antisymmetric tensor fields. (ii) $M$: 16 NS-NS moduli
fields paramaterized by a real symmetric $4\times 4$ 
matrix $M$ and satisfying the condition $M L M = L$,
where $L$ is the $O(4, 4)$ metric. 
These come from the internal components $g_{i j}$ and 
$B_{i j}$ (i,j = 6..9) of the ten dimensional metric 
$g_{M N}$ and antisymmetric tensor $B_{M N}$
respectively. 
(iii) $A_{\bar{\mu}}$: These are the 8 U(1) gauge fields 
in the NS-NS sector coming from the 
$g_{\bar{\mu} i}$ and $B_{\bar{\mu} i}$ componenets of the ten dimensional 
metric and antisymmetric tensor respectively. 
(iv) $\psi$: 8 scalar fields denoted by an eight-dimensional spinor
representation of $SO(4, 4)$
coming from the RR-sector from the $A_i$ 
components of the ten dimensional gauge fields $A_M$
and $C_{i j k}$ components of the third rank 
antisymmetric tensor $C_{MNP}$ in ten dimensions. 
(v) $K_{\bar{\mu}}$: 8 gauge fields coming from the 
RR-sector, 6 of these are the $C_{\bar{\mu} i j}$ components of 
$C_{M N P}$, 1 from $A_{\bar{\mu}}$ component of $A_M$ 
and another 1 from the dualization of the six 
dimensional third rank tensor field 
$C_{\bar{\mu} \bar{\nu} \bar{\rho}}$, 
i.e. $\tilde{C_{\bar{\mu}}}$.
(vi) In addition we have 4 second rank antisymmetric tensor
fields which can also be arranged as a spinor representation
by considering the corresponding 
field strengths as 8 (anti)self-dual
field strengths denoted by 
$D_{\bar{\mu} \bar{\nu} \bar{\rho}}^{\alpha}$.

The fields and transformations with {\it tilde}, namely 
$(\tilde{A}, \tilde{\psi}, \tilde{M}, \tilde{K})$ and 
$\tilde{\Omega}$ are related to those 
without {\it tilde} by a standard
change of basis:
\be
\tilde{L} = \eta^T L \eta
\ee
where
\be
\eta =  {1\over {\sqrt{2}}}
        \left( \begin{array}{cc}
        I_4       &       -I_4      \\
        I_4       &       I_4
\end{array}     \right),
\;\;\;\;\;\;
\tilde{L} = \left( \begin{array}{cc}
        0       &       I_4     \\
        I_4     &       0
\end{array}     \right),
\ee 
and $L$ is the diagonal $O(4, 4)$ metric used in 
\cite{seva}. This further implies
\be
\tilde{\Omega} = \eta^T \Omega \eta,\;\;\; 
\tilde{M} = \eta^T M \eta, \;\;\; etc.
\label{defeta}
\ee
These notations 
do not create any confusion, as in the final spectrum
counting, we directly use the transformations 
(\ref{z2o}) and (\ref{z2*}).
We keep both these notations in order to keep the various
expressions in simple looking forms. 

Unlike in \cite{seva},
We have chosen a more specific form for 
$A_{\bar{\mu}}$, $\psi$, and $K_{\bar{\mu}}$ in equations (\ref{defns}). 
This essentially is a 
choice of basis for arranging the fields in the representaions
of the $SO(4, 4)$ symmetry, and is related to the specific choice
of the eight dimensional $\gamma$-matrices.
A similar choice for the field arrangements has recently been 
presented in ref.\cite{dsen}.
We would also like to emphasize that the explicit choice of field
ararngements in (\ref{defns}) is only to 
use it for understanding the 
discrete symmetries of the theory and should not be taken as any
precise identification among the field contents in the action 
(\ref{action}) and the quantities $A_{\bar{\mu}}$,
$K_{\bar{\mu}}$, $\psi$, defined in \cite{seva}. 
In other words, each of these can have
extra terms which do not change the nature of the discrete
transformation for $A_{\bar{\mu}}$, $\psi$ 
and $K_{\bar{\mu}}$ under the
symmetries (\ref{z2o}) and (\ref{z2*}). 

Now, as pointed out in \cite{seva} and noted earlier 
in this paper, a particular
element of $U$-duality $SO(5, 5)$, namely $U_0$, relates a
dual pair of T-duality symmetries. To show that $U_0$ also
relates the symmetries in equations (\ref{z2o}) 
and (\ref{z2*}), we use the
identification in (\ref{defns}), to put these transformations 
in the same form as
the $T$-duality transformations written in \cite{seva}. 
They are given as:
\bea
M &\rightarrow& \Omega M \Omega^{T},\nonumber\\
\psi^{\alpha} &\rightarrow& \left(
R_{s}\left(\Omega\right)\right)_{\alpha\beta} \psi^{\beta},\nonumber \\
D_{\bar{\mu} \bar{\nu} \bar{\rho}}^{\left( \alpha\right)} &\rightarrow&
\left(R_{s}\left(\Omega\right)\right)_{\alpha \beta}
 D_{\bar{\mu} \bar{\nu} \bar{\rho}}^{\beta},\nonumber\\
A_{\bar{\mu}}^{\left(\alpha\right)} &\rightarrow&
\Omega_{ab}A_{\bar{\mu}}^{\left( b\right)},  \nonumber\\
K_{\bar{\mu}}^{\left( \alpha'\right)} &\rightarrow& 
\left(R_{c}\left(\Omega\right)\right)_{\alpha'\beta'}
K_{\bar{\mu}}^{\left(\beta'\right)}.
                        \label{tdual}
\eea

For the $T$-duality case, other fields, $g_{\bar{\mu}
\bar{\nu}}$, $B_{\bar{\mu}\bar{\nu}}$ and $\phi$ remain
invariant. 
The result of a transformation by $U_0$ is that a T-duality
rotation represented by a matrix $\Omega$ has a dual pair
$\Omega' = R_c(\Omega)$, where $R_c(\Omega)$ is the conjugate
spinor representation for $SO(4, 4)$\cite{seva}. 
For later use we also 
mention that for a special class of 
T-dualtity matrices, $\Omega$, 
represented by $(\theta_L, \phi_L, \theta_R, \phi_R)$, 
and given by the matrix expression:
\be
\Omega = \pmatrix{\omega\left(\theta_{L}\right)&  &  &  \cr
&\omega\left(\phi_{L}\right)&  &  \cr
  & & \omega\left(\theta_{R}\right) &  \cr
   & & &\omega\left(\phi_{R}\right)},
\ee
where
\be
\omega\left(\theta\right)= \pmatrix { \cos \theta & \sin \theta
\cr  -\sin \theta  & \cos \theta },
\ee
the transformation matrices for
spinors, namely $R_s(\Omega)$ and $R_c(\Omega)$, take the 
form:
\be
R_s\left(\Omega\right)=\left(\begin{array}{cccc} \theta_{L}^{''}
,&\phi_{L}^{''},&\theta_{R}^{''} &\phi_{R}^{''}\end{array}\right),
\ee
where,
\be
\pmatrix{\theta_{L}^{''}\cr \phi_{L}{''} \cr \theta_{R}^{''} \cr
\phi_{R}^{''}} =A_{s}\pmatrix{\theta_{L} \cr \phi_{L}\cr
\theta_{R}\cr\phi_{R}}, \;\;
A_{s}=\pmatrix{ 1/2 &1/2 & 1/2 &-1/2 \cr
                1/2 &1/2 &-1/2 &1/2  \cr
                1/2 &-1/2&1/2 &1/2 \cr
               -1/2 &1/2 &1/2 &1/2}.
\ee
Similarly,
\be
R_{c}\left(\Omega\right)=\left(\begin{array}{cccc}
\theta_{L}^{'},&\phi_{L}^{'},\theta_{R}^{'},\phi_{R}^{'}
\end{array}\right),
\ee
where,
\be
\pmatrix{\theta_{L}^{'}\cr\phi_{L}^{'}\cr\theta_{R}^{'}\cr\phi_{R}^{'}}=
A_{c}\pmatrix{\theta_{L}\cr\phi_{L}\cr\theta_{R}\cr\phi_{R}}, \;\;
A_{c}=\pmatrix{1/2 &-1/2 &1/2 &-1/2\cr 
               -1/2&1/2 &1/2 &-1/2\cr
               1/2&1/2&1/2&1/2\cr
              -1/2 &-1/2 &1/2 &1/2}.
\ee 

We now use these notations to rewrite the transformations 
(\ref{z2o}) and
(\ref{z2*}) in the form of equations (\ref{tdual}). We have,
\bea
\tilde{A}_{\bar{\mu}}&\rightarrow& \pmatrix{-I_{4}& \cr
                                  &I_{4}}\tilde{A}_{\bar{\mu}}
\equiv \tilde{\Omega}_{1} \tilde{A}_{\bar{\mu}},  \cr 
\tilde{M} &\rightarrow& \tilde{\Omega}_{1} \tilde{M} 
        \tilde{\Omega}_{1}^{T}, \nonumber \cr
\tilde{\psi} &\rightarrow& \pmatrix{-I_{4}& \cr
                                &I_{4}}\tilde{\psi}
              \equiv \tilde{R}_s(\tilde{\Omega}_1) 
                \tilde{\psi}, \nonumber \\
\label{transform1}
\tilde{K}_{\bar{\mu}}&\rightarrow& \pmatrix{-I_{3}& &  & \cr
                                 &1&  &  \cr
                                 & &-I_{3} & \cr 
                                 & &  &1}\tilde{K}_{\bar{\mu}}
\equiv \tilde{R}_{c}\left(\tilde{\Omega}_{1}\right) 
        \tilde{K}_{\bar{\mu}},
\eea
and
\bea
\tilde{A}_{\bar{\mu}}&\rightarrow& \pmatrix{-I_{3}& & &  \cr
                                    &1& &  \cr
                                    & &-I_{3} &  \cr
                                    & &  &1  }
                        \tilde{A}_{\bar{\mu}}
\equiv \tilde{\Omega}_{2}A_{\bar{\mu}} ,\nonumber\\
\tilde{M}&\rightarrow& \tilde{\Omega}_{2}\tilde{M}
        \tilde{\Omega}_{2}^{T} ,\nonumber\\
\tilde{\psi} &\rightarrow& \pmatrix{ -I_{3}&  &  &  \cr
                                 &1 &  &  \cr
                                 &  &I_{3} &  \cr
                                 &  &  &-1} \psi 
            \equiv \tilde{R}_s(\tilde{\Omega}_2)\psi,\nonumber\\
\tilde{K}_{\bar{\mu}}&\rightarrow& \pmatrix{-I_{4} &  \cr
                           & I_{4}}\tilde{K}_{\bar{\mu}}
         \equiv \tilde{R}_c(\tilde{\Omega}_2)\tilde{K}_{\bar{\mu}}.
        \label{transform2}
\eea
Nontrivial checks on the fact that the symmetries in 
(\ref{transform1}) and
(\ref{transform2}) are connected by $U_0$ 
is provided by the fact that the properties\cite{seva}:
\be
 \tilde{R}_c(\tilde{\Omega}_1) = 
\tilde{\Omega}_2, \;\;\tilde{R}_c(\tilde{\Omega}_2) 
=  \tilde{\Omega}_1, \;\;and \;\;
\tilde{R}_s(\tilde{R}_c(\tilde{\Omega}_1)) 
= \tilde{\Omega}_0 \tilde{R}_s(\tilde{\Omega}_1) 
\tilde{\Omega}_0^{-1}.
                        \label{identity}
\ee
are seen to satisfy by comparing 
$\tilde{R}_s(\tilde{\Omega}_i)$ and 
$\tilde{R}_c(\tilde{\Omega}_i)$ in equations (\ref{transform1}) 
and (\ref{transform2}), where 
$\Omega_0$ is given by the matrix:
\be
\Omega_0 = \pmatrix{I_2 & &  \cr
                        &\sigma_{3} & \cr
                        &   & I_4 }.
\ee

Given $\Omega_1$ and $\Omega_2$, 
the form of $R_s(\Omega)$ and $R_c(\Omega)$ depends on the
choice of eight dimensional Dirac $\gamma$ matrices. Our field 
arrangements in equation (\ref{defns})
provide one such choice.
The fact that $R_s(\Omega_i)$ and $R_c(\Omega_i)$ 
of equations (\ref{transform1}) and 
(\ref{transform2}) are related by
relationships in (\ref{identity}) 
confirms that we have made a consistent
choice. As mentioned earlier, another choice for field 
arrangements has been presented in a different context in 
\cite{dsen}.

Another way to notice that we have the correct dual pair of
symmetries, $Z_2^o$ and $Z_2^*$,
is to compare our results with those for the $K3$
compactification in \cite{vawit},
where it was argued that the dual 
of a $B \rightarrow - B$ symmetry lies in the center of the 
resulting moduli space group, when $B$ is set to zero. We notice
that our $\tilde{\Omega}_2 \equiv 
\tilde{R}_c(\tilde{\Omega}_1)$ satisfies this
property, as it lies in the center of the group $SO(3, 3) \times
SO(1, 1)$, which by using the equivalence, $SL(4)\sim SO(3, 3)$,
turns out to be the group of constant 
coordinate transformations on four of the compactified
coordinates. 

We now present the construction of two sets of examples of 
dual pairs in four dimensions following the
methods in \cite{seva}.
To construct a four dimensional example,
one starts from
the six dimensional theory and further compactifies on the 
two-torus. We also mod out the original spectrum by the
dual pair of discrete symmetries. This means that 
only those 
states are kept which are invariant under the  
discrete symmetries. To avoid any states coming from
the twisted sectors, one combines these discrete 
symmetries with the shifts along the central charges of the
sypersymmetry algebra. The simplest of these being the 
translations along one of the two compactified directions 
which we choose, in this case, as $x^4$. 
We now construct a dual pair of $N=4$ supersymmetric theories 
in four dimensions
by modding out the six dimensional theories by the discrete 
symmetries in (\ref{z2o})-(\ref{z2*}) or 
(\ref{transform1})-(\ref{transform2}) and 
compactifying two extra dimensions. 
The states which are invariant under  the $Z_2^o$ 
symmetry of (\ref{z2o}), together with an operation
which shifts $x^4$ by half its periodicity, are given
as:

(i) Vectors: $g_{\hat{\mu} 4}$, $A_{\hat{\mu}}$, 
$B_{\hat{\mu} i}$,
$B_{\hat{\mu} 5}$, $C_{\hat{\mu} 4 i}$ and 
$C_{\hat{\mu} 4 5}$, where as mentioned before, $\hat{\mu}$ is
now a four dimensional vector index. 
Total number of these fields is therefore 12.  Since the 
resulting theory, as argued below,
is an $N=4$ supersymmetric theory, the 
number of vector fields completely determines 
the local moduli space of scalars. In this case 
it should be given by the coset 
${SL(2, R)\over U(1)} \times {O(6,6)\over {O(6)\times O(6)}}$.
The total number of scalar fields surviving the above 
projection should therefore be 38. In the following we 
verify this by direct counting.

(ii) Scalars: $B_{45}$, $B_{4i}$, $g_{44}$,
$g_{ij}$, $g_{55}$, $g_{i5}$, $A_4$, $C_{ijk}$
and $C_{ij5}$. These are 32 scalar fields.
We however get additional scalar fields by dualizing 
the surviving antisymmetric tensor fields in four dimensions. 

(iii) Antisymmetric tensors: $C_{\hat{\mu} \hat{\nu} i}$ and 
$C_{\hat{\mu} \hat{\nu} 5}$. These are 
five antisymmetric tensors. 
They give additional 5 scalar fields upon dualization.

(iv) The rest of the surviving degrees of freedom are 
the graviton $g_{\mu \nu}$ and the dilaton $\phi$.

As a result we see that all the scalars coming from (ii),
(iii) together with dilaton give the right number of 
scalar fields needed from the supersymmetry requirements.
The spectrum of the dual model is found by modding out the 
six dimensional theory by the $Z_2^*$ symmetry  
(\ref{z2*}) together with a half shift on $x^4$. The states 
surviving the projections are:

(i$'$) vectors: $g_{\hat{\mu} 4}$, 
$g_{\hat{\mu} 9}$, $B_{\hat{\mu} 4}$,
$B_{\hat{\mu} 9}$, $A_{\hat{\mu}}$, 
$C_{\hat{\mu} 4 9}$, $C_{\hat{\mu} m 5}$
and $C_{\mu m n}$, where $m,n \equiv 6, 7, 8$. 
Once again we have 12 gauge fields which
shows the matching of the vector field spectrum under the 
modding out operation by the two symmetries 
of equations (\ref{z2o}) and (\ref{z2*}). 

(ii$'$) Scalars: $g_{44}$, $g_{m n}$, $g_{m 5}$, $g_{5 5}$,
$g_{4 9}$, $g_{9 9}$, $B_{m n}$, $B_{m 5}$, $B_{4 9}$,
$A_4$, $A_9$, $C_{5 9 m}$, $C_{9 m_1 m_2}$, 
$C_{4 5 m}$, $C_{4 m_1 m_2}$. These are 34 scalar 
fields. Additional scalar fields once again come by the
dualization of the antisymmetric tensors.

(iii$'$) Antisymmetric Tensors: 
$C_{\hat{\mu}_1 \hat{\mu}_2 4}$,
$C_{\hat{\mu}_1 \hat{\mu}_2 9}$ and 
$B_{\hat{\mu} \hat{\nu}}$. 

(iv$'$) In addition we once again have the graviton and 
the dilaton.

All the 34 scalars in (ii), together with the ones 
coming from the the dualization of the 3 antisymmetric 
tensors and the dilaton provide once again the 38
scalars needed to make the proper moduli space for
the $N=4$ supersymmetric theory, as well as 
the number required by the string-string duality.

The number of supersymmetries in the two cases also match.
In the first case the number of supersymmetries is 
reduced by a factor of half for the simple reason that 
one identifies the states in the left-moving part 
with the ones in the right-moving ones. In the 
dual case also, the supersymmetries reduce by a factor of
half due to the appropriate projections. In this case,
by looking at the form of $\tilde{\Omega_2}$
in (\ref{transform2}),
which from equation (\ref{defeta}) is also equal to 
$\Omega_2$, we find that it 
corresponds to an additional left-right symmetric
GSO projection. As a result we get a $(2, 2)$ 
space-time supersymmetric model in this case. 

We have thus presented the results
of the projection with respect to the 
symmetries discovered in equations 
(\ref{z2o}) and (\ref{z2*}) of this paper. It implies that 
the weak coupling limit of a type II theory defined
on a non-orientable worldsheet, called orientifold, is equivalent
in the strong coupling limit,   to another string theory with 
$(2, 2)$ space-time supersymmetry on an orientable 
worldsheet and vice versa.

We can also construct new dual pair of 
models by combining 
the symmetries $Z_2^o$ and $Z_2^*$ with the T-duality 
symmetries of \cite{seva}. As an explicit case, we consider the 
six dimensional example presented in \cite{seva}
and its further compactification to four dimensions.  
The first $N=2$ 
model is now constructed from $N=4$ one 
in six dimensions by modding it out by a $Z_2^1
\times Z_2^o$ symmetry. The first of this $Z_2$, i.e. $Z_2^1$, 
corresponds to 
$(\theta'_L, \phi'_L, \theta'_R, \phi'_R) = 
(\pi, - \pi, \pi, -\pi)$, together with an RR gauge
transformation, to keep the twisted sector heavy. Note that the
original $N=4$ theory as well as the modding out procedure 
by $Z_2^1$ is 
left-right symmetric on the worldsheet. As a result this allows
us a further modding out by $Z_2^o$. 
The product of these two $Z_2$'s is a valid projection, since they
commute with each other. This is further confirmed by the
counting of the massless spectrum after these projections 
and the projections which are dual to these. 

The mass spectrum corresponding to 
$Z_2^1 \times Z_2^0$ projections are found
by taking the intersections of the states 
discussed in (i)-(iv) above with
those in \cite{seva} for the six dimensional example. The surviving
states are:
(i) vectors: $g_{\hat{\mu} 4}$, $B_{\hat{\mu} 5}$, 
$A_{\hat{\mu}}$ and 
$C_{\hat{\mu} 4 5}$. This theory therefore has four 
$U(1)$ gauge fields.  
(ii) Scalars: $g_{i j}$, $g_{4 4}$, $g_{5 5 }$, $B_{4 5}$, 
$A_4$, $C_{i j 5}$.
(iii) In addition we have the dilaton, the antisymmetric tensor
$C_{\hat{\mu} \hat{\nu} 5}$ and the graviton 
$g_{\hat{\mu} \hat{\nu}}$. 
As a result we have total 22 scalar fields when we combine the
scalars coming from (ii) with the dilaton and the dual of
$C_{\hat{\mu} \hat{\nu} 5}$. 

To obtain the dual theory, we have to mod out the orignal $N=4$
theory by another product of $Z_2$'s, namely $Z_2^2 \times
Z_2^*$.  $Z_2^2$ is the T-duality
symmetry represented by $(\theta_L, \phi_L, \theta_R, \phi_R) =
(2\pi, 0, 0, 0)$ \cite{seva} together with a left-right
symmetric half lattice shift in the $\phi$ plane.
$Z_2^*$ is the improper T-duality
rotation of equation (\ref{z2*}). 
The final result of these projections
gives the following vector and scalar degrees of freedom. 
(i) Vectors: $g_{\hat{\mu} 4}$, $B_{\hat{\mu} 4}$, 
$g_{\hat{\mu} 9}$, $B_{\hat{\mu} 9}$.
As a result we once again have four surviving $U(1)$ vector
fields as required by the duality symmetry.
(ii) Scalars: $B_{{m} {n}}$, $B_{{m}, 5}$,
$B_{4 9}$, $g_{4 4}$, $g_{{m} {n}}$, 
$g_{{m} 5}$, $g_{5 5}$, $g_{4 9}$, $g_{9 9}$.
(iii) In addition we have the graviton 
$g_{\hat{\mu} \hat{\nu}}$, dilaton
$\phi$ and the antisymmetric tensor 
$B_{\hat{\mu} \hat{\nu}}$. Once again
the number of all the scalars in (ii) together with dilaton and
the dual of the antisymmetric tensor gives the right number 22
for the sclar fields. 
Both these models are examples of $N=2$ supersymmetric theories
in 4-dimensions. The first one gives an $N=2$
supersymmetric model where the type II theory is described on an
orientifold. On the other hand its dual pair is a
string theory on normal worldsheet where supersymmetry
contributions from the left and right sectors are (1, 1). 

To conclude, 
We have shown that the projections with respect to the new
discrete symmetries presented in equations (\ref{z2o}) 
and (\ref{z2*}) lead to a
consistent dual pair of theories. We have also combined $Z_2^o$
and $Z_2^*$ with many other discrete symmetries of \cite{seva} and
found the matching of the spectrum. 
The details of these results, together
with the status of symmetries in these models, will be presented
separately.

\section*{Acknowledgement} 

I thank Ashoke Sen for many invaluable suggestions and
discussions. 
I thank Anindya Biswas for verifying many results of
this paper and for many other useful discussions. I also thank
the Mehta Research Institute, Allahabad for its hospitality,
where part of this work was done. 
 


\newpage
\begin{thebibliography}{99}
\newcommand{\np}{{\it Nucl. Phys.} {\bf B}}
\newcommand{\pl}{{\it Phys. Lett.} {\bf B}}
\newcommand{\prd}{{\it Phy. Rev. }{\bf D}}
\newcommand{\prl}{{\it Phys. Rev. Lett.}}
\newcommand{\mpl}{{\it Mod. Phys. Lett. }{\bf A}}
\newcommand{\ijmp}{{\it Int. J. Mod. Phys.}{\bf A}}
\newcommand{\cqg}{\it Class. Quant. Grav.}
\newcommand{\jmp}{\it J. Math Phys.}
\newcommand{\cmp}{\it Comm. Math. Phys.}

\bibitem{duff} M. Duff, ``Strong/Weak Coupling Duality from the 
Dual Strings", Newton Institute Preprint,
hep-th/9501030. 
\bibitem{hull} C. Hull and P. Townsend, ``Unity of Superstring 
Dualities", hep-th/9510167.
\bibitem{witten} E. Witten, ``String Theory Dynamics in Various
Dimensions", {\np} {\bf 443} (1995) 85.
\bibitem{townsend} C. Hull and P. Townsend, ``Enhanced Gauge 
Symmetry in Superstring Theory", hep-th/9505073; P. Aspinwall,
``An N=2 Dual Pair and a Phase Transition", hep-th/9510142;
``Dualities in Five Dimensions and Charged String Solutions", 
S. Kar, J. Maharana and S. Panda, hep-th/9511213.
\bibitem{sen1} A. Sen, ``String-String Duality Conjecture in 
Six Dimensions and Charged Solitonic Strings", {\np} {\bf 450}
(1995) 103.
\bibitem{harvey} J. Harvey and A. Strominger, ``The Heterotic
String as a Soliton", {\np} {\bf 449} (1995) 535.
\bibitem{ferrara} S. Ferrara, J. Harvey, A. Strominger and C.
Vafa, ``Second Quantized Mirror Symmetry", hep-th/9505162.
\bibitem{kachru} S. Kachru and C. Vafa, ``Exact Results for 
N=2 Compactification of Heterotic String", {\np} {\bf 450} 
(1995) 69. 
\bibitem{anton} I. Antoniadis, S. Ferrara and T. Taylor, 
``N=2 Heterotic Superstring and its Dual Theory in Five 
Dimensions", hep-th/9511116; I. Antoniadis and H. Partouche,
``Exact Monodromy Group of N=2 Heterotic Superstring", 
hep-th/9509009.
\bibitem{vawit} C. Vafa and E. Witten, ``Dual String Pairs with 
N=1 and N=2 Supersymmetry in Four Dimensions", {\np}{\bf} (1995)
, hep-th/9507050.
\bibitem{scsen} J. Schwarz and A. Sen, ``The type IIA dual of the
six Dimesnsional CHL compactification", hep-th/9507027.
\bibitem{chaudh} S. Chaudhuri and D. Lowe, ``Type IIA-Heterotic
Duals with Maximal Supersymmetry", hep-th/9508144.
\bibitem{quev} G. Aldazabal, L. Ibanez, A. Font and 
F. Quevedo, ``Chains of N=2, D=4 heterotic/typeII duals",
hep-th/9510093.
\bibitem{gao} H. Gao, ``More Dual Pairs from Orbifolding",
hep-th/9512060. 
\bibitem{seva} A. Sen and C. Vafa, ``Dual Pairs of Type II String
Compactification", hep-th/9508064.
\bibitem{horava} A. Sagnotti, ``Open Strings and their Symmetry
Groups", in {\it Non-perturbative Quantum Field Theory}, 
Cargese 1987, eds. G. Mack et. al. (Pergamon Press 1988);
P. Horava, ``String on Worldsheet Orbifolds",
{\np} {\bf 327} (1989) 461; ``Background Duality of Open String
Models", {\pl} {\bf 321} (1989) 251.
\bibitem{leigh} J. Dai, R. Leigh, and J. Polchinski, ``New
Connections between String Theories", {\mpl} {\bf A4} (1989)
2073. 
\bibitem{mukhi} P. Horava and E. Witten, ``Heterotic and Type I
String Dynamics from Eleven Dimensions", hep-th/9510209; 
K. Dasgupta and S. Mukhi, ``Orbifolds of M-theory",
hep-th/9512196; E. Witten, ``Five Branes and M-Theory on on 
Orbifold", hep-th/9512219.
\bibitem{dsen} A. Sen, ``U-Duality and Intersecting D-Branes",
hep-th/9511026. 
\end{thebibliography}
\end{document}